\documentclass[a4paper]{PoS}

\usepackage{amsmath}
\usepackage{amssymb}
\usepackage{amsfonts}
\usepackage{amsthm}

\newcommand{\ee}[0]{\mathrm{e}}
\newcommand{\ii}[0]{\mathrm{i}}

\newcommand{\identity}[0]{\mathbb{I}}

\newcommand{\vect}[1]{\mathbf{#1}}
\newcommand{\unitvect}[1]{\hat{\mathbf{#1}}}

\newcommand{\adjoint}[1]{\smash{\overline{#1}}\vphantom{#1}}

\newcommand{\transpose}{^\top}

\DeclareMathOperator{\tr}{tr}

\newcommand{\definedby}{:=\,}

\newtheorem{proposition}{Proposition}[section]

\usepackage{braket}
\usepackage{siunitx}
\usepackage{nicefrac}

\newcommand{\projector}[1]{\Gamma_{\!#1}}
\newcommand{\proj}[1]{\projector{\vect{#1}}}
\newcommand{\spinor}[1]{u_{#1}(p, s)}

\title{Electromagnetic Form Factors of Excited Nucleons via Parity-Expanded Variational Analysis}

\ShortTitle{Electromagnetic Form Factors of Excited Nucleons}

\author{\speaker{Finn M. Stokes}, Waseem Kamleh, Derek B. Leinweber and Benjamin J. Owen\\
        Centre for the Subatomic Structure of Matter,\\
	Department of Physics,\\
        University of Adelaide, SA 5005\\
	E-mail: \email{finn.stokes@adelaide.edu.au}}

\abstract{Variational analysis techniques in lattice QCD are powerful tools that give access
to the excited state spectrum of QCD\@. At zero momentum, these techniques are well established and
can cleanly isolate energy eigenstates of either positive or negative parity.
In order to compute the form factors of a single energy eigenstate, we must perform a
variational analysis at non-zero momentum. When we do this with baryons, we run into issues
with parity mixing, as boosted baryons are not eigenstates of parity. The parity-expanded
variational analysis (PEVA) technique is a novel method for ensuring the successful
and consistent isolation of boosted baryon eigenstates. This is achieved through a parity expansion of
the operator basis used to construct the correlation matrix.
World-first calculations of excited state nucleon form factors using this new technique are
presented, showing the improvement over conventional methods.}

\FullConference{34th annual International Symposium on Lattice Field Theory\\
                24--30 July 2016\\
                University of Southampton, UK}

\begin{document}

\section{Introduction}
In order to evaluate the form factors and transition moments of baryon excitations in lattice
QCD, it is necessary to isolate these states at finite momentum. Excited
baryons have been isolated on the lattice through a combination of parity projection and
variational analysis techniques \cite{Kiratidis:2015vpa, Mahbub:2013ala, Mahbub:2012ri, Mahbub:2010rm,
Edwards:2012fx, Edwards:2011jj, Lang:2016hnn, Lang:2012db, Alexandrou:2014mka}.
At zero momentum, these techniques are well
established and can isolate the states of interest. However, at non-zero momentum, these
techniques are vulnerable to opposite parity contaminations.

To resolve this issue, we developed the Parity-Expanded Variational Analysis (PEVA) technique
\cite{Menadue:2013kfi}. By introducing a novel Dirac projector and expanding the operator basis used to
construct the correlation matrix, we are able to cleanly isolate states of both parities at
finite momentum.

Utilising the PEVA technique, we are able to present here the world's first lattice QCD calculations of
nucleon excited state form factors free from opposite parity contaminations. Specifically,
the Sachs electromagnetic form factors of a localised negative parity nucleon
excitation are examined.
Furthermore, we clearly demonstrate the efficacy of variational analysis techniques at providing access
to ground state form factors with extremely good control over excited state effects.

\section{Conventional Variational Analysis}
We begin by briefly highlighting where opposite parity contaminations enter into conventional
variational analysis techniques and motivate the PEVA technique. In these proceedings we use the Pauli
representation for Dirac matrices.
In order to discuss opposite parity contaminations, we need to be able to categorise states by
their parity. However, eigenstates of non-zero momentum are not eigenstates of parity, so we must
categorise boosted states by their rest-frame parity. 

To perform a conventional variational analysis on spin-1/2 baryons, we take a basis of \(n\)
conventional baryon operators \({\chi^i}\),
which couple to states of both parities,
\begin{align*}
        \braket{\Omega | \chi^i | B^+} = \lambda^i_{B^+} \sqrt{\frac{m_{B^+}}{E_{B^+}}} \, \spinor{B^+} \, , \quad
        \braket{\Omega | \chi^i | B^-} = \lambda^i_{B^-} \sqrt{\frac{m_{B^-}}{E_{B^-}}} \, \gamma_5 \, \spinor{B^-}\,,
\end{align*}
and use this basis to form an \(n\times n\) matrix of two-point correlation functions
\begin{align*}
	\mathcal{G}^{ij}(\vect{p};\, t) \definedby &\sum_{\vect{x}} \ee^{\ii\vect{p}\cdot\vect{x}}\, \braket{\Omega | \chi^{i}(x) \, \adjoint\chi^{j}(0) | \Omega} \,.
\end{align*}

This correlation matrix contains states of both parities, so we introduce the
`parity projector' \(\projector{\pm} = \left(\gamma_4\pm\mathbb{I}\right)/2\), and take the spinor trace,
defining the projected correlation matrix
\(G^{ij}(\projector{\pm};\, \vect{p},\, t) \definedby \tr\left(\projector{\pm} \, \mathcal{G}^{ij}(\vect{p},\, t)\right)\).
By inserting a complete set of states
between the
two operators, and noting the use of Euclidean time,
we can rewrite this projected correlation matrix as
\begin{align*}
	G^{ij}(\projector{\pm};\, \vect{p},\, t) 
	=\, & \sum_{B^{+}} \ee^{-E_{B^{+}}(\vect{p})\,t} \, \lambda^i_{B^{+}} \, \adjoint\lambda^j_{B^{+}} \frac{E_{B^{+}}\!(\vect{p}) \,\pm m_{B^{+}}}{2E_{B^{+}}\!(\vect{p})}
	+ \sum_{B^{-}} \ee^{-E_{B^{-}}(\vect{p})\,t} \, \lambda^i_{B^{-}} \, \adjoint\lambda^j_{B^{-}} \frac{E_{B^{-}}\!(\vect{p}) \,\mp m_{B^{-}}}{2E_{B^{-}}\!(\vect{p})}\,,
\end{align*}

At zero momentum, \(E_{B}(\vect{0}) = m_{B}\) and the projected correlation matrices will each contain
terms of a single parity.
However, at non-zero momentum \(E_{B}(\vect{p}) \ne m_{B}\) and the projected correlation matrices
contain \(O\left((E - m)/2E\right)\) opposite parity contaminations.
These opposite parity contaminations were investigated in Ref.~\cite{Lee:1998cx}.

\section{Parity-Expanded Variational Analysis}
\label{sec:PEVA}
To solve the problem of opposite parity contaminations at finite momentum, we developed the PEVA
technique \cite{Menadue:2013kfi}. In this section, we summarise the PEVA technique, and describe how it
applies to form factor calculations.

The PEVA technique works by expanding the operator basis of the correlation matrix to isolate energy
eigenstates of both rest-frame parities simultaneously while still retaining a signature of this parity.
By considering the Dirac structure of the unprojected correlation matrix, we construct the novel
momentum-dependent projector \(\proj{p} \definedby \frac{1}{4} \left(\identity + \gamma_4\right) \left(\identity -
\ii \gamma_5 \gamma_k \unitvect{p}_k\right)\). This allows us to construct a set of ``parity-signature''
projected operators \(\left\{\chi^i_{\vect{p}} = \proj{p} \, \chi^i \, , \; \chi^{i'}_{\vect{p}} =
\proj{p} \, \gamma_5 \, \chi^i\right\}\), where the primed indices denote the inclusion of \(\gamma_5\),
inverting the way the operators transform under parity.
Unlike the conventional opertors \({\,\chi^i\,}\), the inclusion of \(\proj{p}\) ensures that the operators
\(\chi^i_{\vect{p}}\) and \(\chi^{i'}_{\vect{p}}\) have definite parity at zero momentum without requiring
projection by \(\Gamma_{\pm}\).

By performing a variational analysis with this expanded basis \cite{Menadue:2013kfi}, we construct optimised operators \(\phi^{\alpha}_{\vect{p}}(x)\)
that couple to each state \(\alpha\). We can then use these operators to calculate the three point
correlation function
\begin{align*}
	\mathcal{G}^{\mu}_{+}(\vect{p'}, \vect{p};\, t_2, t_1;\, \alpha) \definedby &\sum_{\vect{x_2},\vect{x_1}} e^{-\ii \vect{p'}\cdot\vect{x_2}} \, e^{\ii (\vect{p'} - \vect{p})\cdot\vect{x_1}} \braket{\,\phi^{\alpha}_{\vect{p'}}(x_2)\,|\,J^{\mu}(x_1)\,|\,\adjoint\phi^{\alpha}_{\vect{p}}(0)\,}\,,
\end{align*}
where \(J^{\mu}\) is the \(O(a)\)-improved \cite{Martinelli:1990ny} conserved vector current used in
\cite{Boinepalli:2006xd}, inserted with some momentum transfer \(q = p' - p\).
We can take the spinor trace of this with some projector \(\Gamma\) to get the projected three point correlation function
\(G^{\mu}_{+}(\vect{p'}, \vect{p};\, t_2, t_1;\, \Gamma;\, \alpha) \definedby \tr\left(\Gamma \, \mathcal{G}^{\mu}_{+}(\vect{p'}, \vect{p};\, t_2, t_1;\, \alpha) \right)\).

There is an arbitrary sign choice in the definition of \(\proj{p}\),
so it is convenient to define \(\proj{p}' \definedby \frac{1}{4}
\left(\identity + \gamma_4\right) \left(\identity + \ii \gamma_5 \gamma_k \unitvect{p}_k\right) =
\proj{-p}\), which is equally valid. We can then use this alternate projector in constructing an
alternate sink operator \(\phi'^{\alpha}_{\vect{p}}(x)\), while leaving the source operator unchanged.
This gives us an alternate three point correlation function, 
\(\mathcal{G}^{\mu}_{-}(\vect{p'}, \vect{p};\, t_2, t_1;\, \alpha)\),
leading to an alternate projected three point correlation function,
\(G^{\mu}_{-}(\vect{p'}, \vect{p};\, t_2, t_1;\, \Gamma;\, \alpha)\).

We can then construct the reduced ratio,
\begin{align*}
	\adjoint{R}_{\pm}(\vect{p'}, \vect{p};\, \alpha;\, r, s) \definedby \, & \sqrt{\left|\frac{r_{\mu} \, G^{\mu}_{\pm}(\vect{p'}, \vect{p};\, t_2, t_1;\, s_{\nu} \, \projector{\nu};\, \alpha) \; r_{\rho} \, G^{\rho}_{\pm}(\vect{p}, \vect{p'};\, t_2, t_1;\, s_{\sigma} \, \projector{\sigma};\, \alpha)}{G(\vect{p'};\, t_2;\, \alpha) \, G(\vect{p};\, t_2;\, \alpha)}\right|} \\
	& \quad \times \mathrm{sign}\left(r_{\gamma} \, G^{\gamma}_{\pm}(\vect{p'}, \vect{p};\, t_2, t_1;\, s_{\delta}\, \Gamma_{\delta};\, \alpha)\right) \sqrt{\frac{2 E_{\alpha}(\vect{p})}{E_{\alpha}(\vect{p})+m_{\alpha}}} \, \sqrt{\frac{2 E_{\alpha}(\vect{p'})}{E_{\alpha}(\vect{p'})+m_{\alpha}}}\,,
\end{align*}
where \(r_{\mu}\) and \(s_{\mu}\) are coefficients selected to determine the form factors. By
investigating the \(r_{\mu}\) and \(s_{\mu}\) dependence of \(\adjoint{R}_{\pm}\), we find that the
clearest signals are given by
\begin{align*}
	R^{T}_{\pm} &= \frac{2}{1 \pm\, \unitvect{p} \cdot \unitvect{p'}} \; \adjoint{R}_{\pm}\left(\vect{p'}, \vect{p};\, \alpha;\, (1, \vect{0}), (1, \vect{0})\right) \,,\;\mathrm{and} \\
	R^{S}_{\mp} &= \frac{2}{1 \pm\, \unitvect{p} \cdot \unitvect{p'}} \; \adjoint{R}_{\mp}\left(\vect{p'}, \vect{p};\, \alpha;\, (0, \unitvect{r}), (0, \unitvect{s})\right) \,,
\end{align*}
where \(\unitvect{s}\) is chosen such that \(\vect{p} \cdot \unitvect{s} = 0 = \vect{p'} \cdot \unitvect{s}\),
\(\unitvect{r}\) is equal to \(\unitvect{q} \times \unitvect{s}\), and the sign \(\pm\) is chosen
such that \(1 \pm \unitvect{p} \cdot \unitvect{p'}\) is maximised.

We can then find the Sachs electric and magnetic form factors
\begin{align*}
	G_E(Q^2) =\, &\left[Q^2 \left(E_{\alpha}(\vect{p'}) + E_{\alpha}(\vect{p})\right)\, \left(\left(E_{\alpha}(\vect{p}) + m_{\alpha}\right)\,\left(E_{\alpha}(\vect{p'}) + m_{\alpha}\right) \mp \bigl|\vect{p}\bigr| \bigl|\vect{p'}\bigr|\right) \, R^{T}_\pm \right. \\
	& \left. \quad \pm\, 2 \bigl|\vect{q}\bigr| \left(1 \mp\, \unitvect{p} \cdot \unitvect{p'}\right) \bigl|\vect{p}\bigr| \bigl|\vect{p'}\bigr| \left(\left(E_{\alpha}(\vect{p}) + m_{\alpha}\right)\,\left(E_{\alpha}(\vect{p'}) + m_{\alpha}\right) \pm \bigl|\vect{p}\bigr| \bigl|\vect{p'}\bigr|\right) \, R^{S}_{\mp} \right] \\
	&/ \left[ 4 m_{\alpha} \left[ \left( E_{\alpha}(\vect{p}) \, E_{\alpha}(\vect{p'}) + m_{\alpha}^2 \mp\, \bigl|\vect{p}\bigr| \bigl|\vect{p'}\bigr| \right) \bigl|\vect{q}\bigr|^2 + 4 \bigl|\vect{p}\bigr|^2 \bigl|\vect{p'}\bigr|^2 \left(1 \mp\, \unitvect{p} \cdot \unitvect{p'}\right) \right] \right] \,,\;\mathrm{and} \\
	G_M(Q^2) =\, &\left[\pm\, 2 \left(1 \mp\, \unitvect{p} \cdot \unitvect{p'}\right) \bigl|\vect{p}\bigr| \bigl|\vect{p'}\bigr| \left(\left(E_{\alpha}(\vect{p}) + m_{\alpha}\right)\,\left(E_{\alpha}(\vect{p'}) + m_{\alpha}\right) \pm \bigl|\vect{p}\bigr| \bigl|\vect{p'}\bigr|\right) \, R^{T}_\pm \right. \\
	& \left. \quad - \bigl|\vect{q}\bigr| \left(E_{\alpha}(\vect{p'}) + E_{\alpha}(\vect{p})\right)\, \left(\left(E_{\alpha}(\vect{p}) + m_{\alpha}\right)\,\left(E_{\alpha}(\vect{p'}) + m_{\alpha}\right) \mp \bigl|\vect{p}\bigr| \bigl|\vect{p'}\bigr|\right) \, R^{S}_{\mp} \right] \\
	&/ \left[ 2 \left[ \left( E_{\alpha}(\vect{p}) \, E_{\alpha}(\vect{p'}) + m_{\alpha}^2 \mp\, \bigl|\vect{p}\bigr| \bigl|\vect{p'}\bigr| \right) \bigl|\vect{q}\bigr|^2 + 4 \bigl|\vect{p}\bigr|^2 \bigl|\vect{p'}\bigr|^2 \left(1 \mp\, \unitvect{p} \cdot \unitvect{p'}\right) \right] \right]\,.
\end{align*}

The details of this procedure will be presented in full in Ref.~\cite{stokes:formfactors}.

\section{Results}

In this section, we present world-first lattice QCD calculations of the Sachs electric and magnetic form factors of the
ground state nucleon and first negative-parity excitation with good control over opposite
parity contaminations. We compare the results obtained by the PEVA technique
to an analysis using conventional parity projection.

These results are calculated on the second
heaviest PACS-CS \((2+1)\)-flavour full-QCD ensemble \cite{Aoki:2008sm}, made available through the ILDG
\cite{Beckett:2009cb}. This ensemble uses a \(32^3 \times 64\) lattice, and employs an
Iwasaki gauge action with \(\beta = 1.90\) and non-perturbatively \(O(a)\)-improved Wilson quarks. We use the
\(m_{\pi}=\SI{570}{\mega\electronvolt}\) PACS-CS ensemble, and set the scale using the Sommer parameter with
\(r_0 = \SI{0.49}{\femto\meter}\), giving a lattice spacing of \(a = \SI{0.1009(23)}{\femto\meter}\). With this scale,
our pion mass is \SI{515(8)}{\mega\electronvolt}. We used 343 gauge field configurations,
with a single source location on each configuration. \(\chi^2 / \mathrm{dof}\) is calculared with the
full covariance matrix, and all fits have \(\chi^2 / \mathrm{dof}\ < 1.2\).

For the analyses in this section, we start with a basis of eight operators, by taking two conventional
spin-\(\nicefrac{1}{2}\) nucleon operators (\(\,\chi_1 = \epsilon^{abc} \, [{u^a}\transpose (C\gamma_5) \, d^b] \, u^c\),
and \(\chi_2 = \epsilon^{abc} \, [{u^a}\transpose (C) \, d^b] \, \gamma_5 \, u^c\,\)),
and applying 16, 35, 100, and 200 sweeps of gauge invariant Gaussian smearing when creating the propagators \cite{Mahbub:2013ala}.
For the conventional variational analysis, we take this basis of eight operators and project with \(\projector{\pm}\),
and for the PEVA analysis, we parity expand the basis to sixteen operators and project with \(\proj{p}\) and \(\proj{p}'\).

To extract the form factors, we fix the source at time slice 16, and the current insertion at time slice 21.
We choose time slice 21 by inspecting the two point correlation functions associated with each state
and observing that excited state contaminations
are strongly suppressed by time slice 21. We then extract the form factors from the ratios given in
Sec.~\ref{sec:PEVA} for every possible sink time and look for a plateau consistent with a single-state
ansatz.

\begin{figure}[t]
	\begin{center}
		\includegraphics[width=0.8\textwidth, trim={0 15mm 0 3mm}]{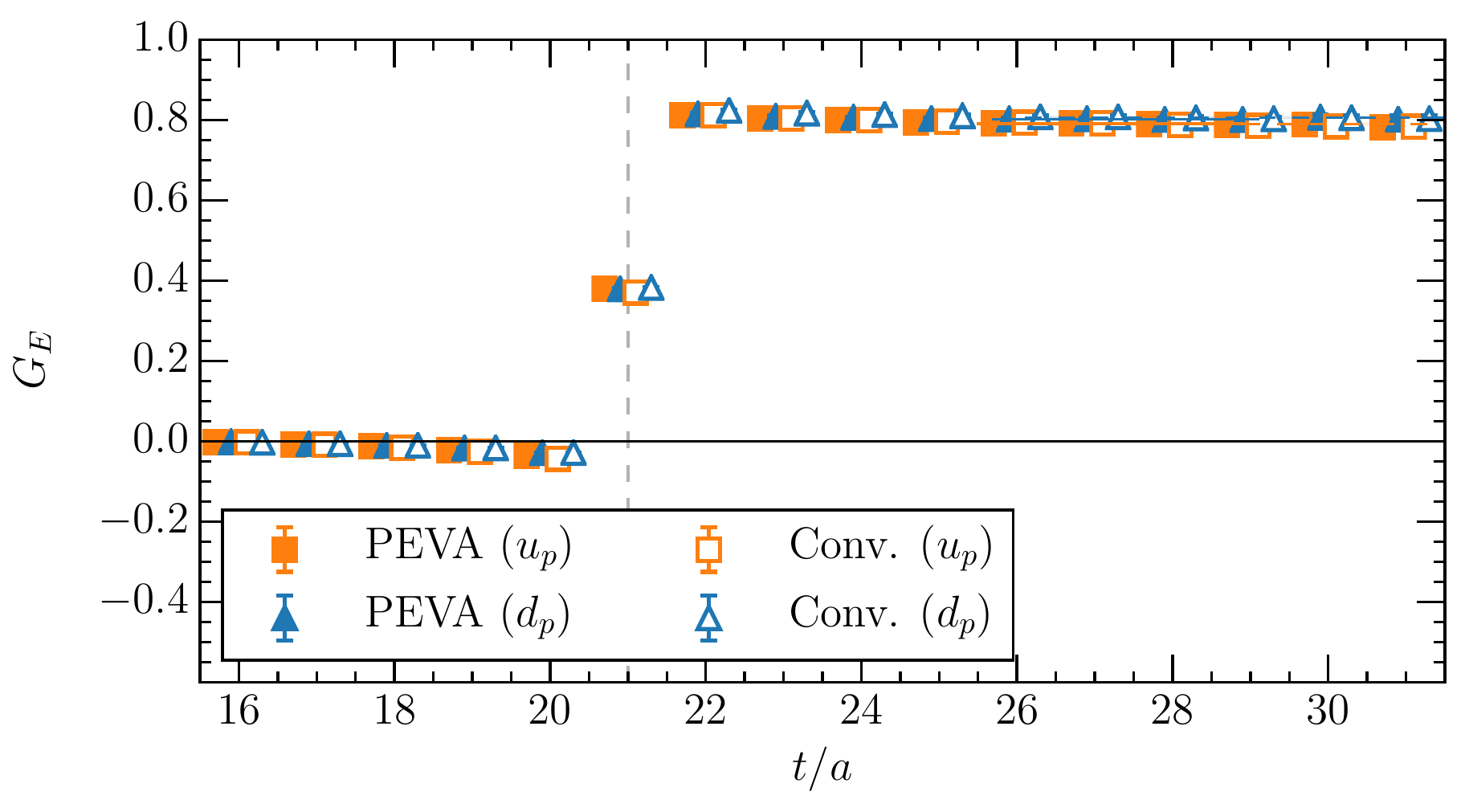}
	\end{center}
	\caption{\label{fig:groundstate:GE}\(G_E(Q^2)\) for the ground state nucleon at
        \(Q^2 = \SI{0.144}{\giga\electronvolt^2}\).
	We plot the conventional analysis with open markers and the new PEVA analysis with closed
	markers. Our fits to the plateaus are illustrated by shaded bands, with the central value indicated by
	dashed lines for the conventional analysis, and solid lines for the PEVA analysis. We plot the
	contributions from the singly represented quark sector
	with blue triangles, and doubly represented quark sector
	with orange squares, for single quarks of unit charge.}
\end{figure}

\begin{figure}[t]
	\begin{center}
		\includegraphics[width=0.8\textwidth, trim={0 15mm 0 3mm}]{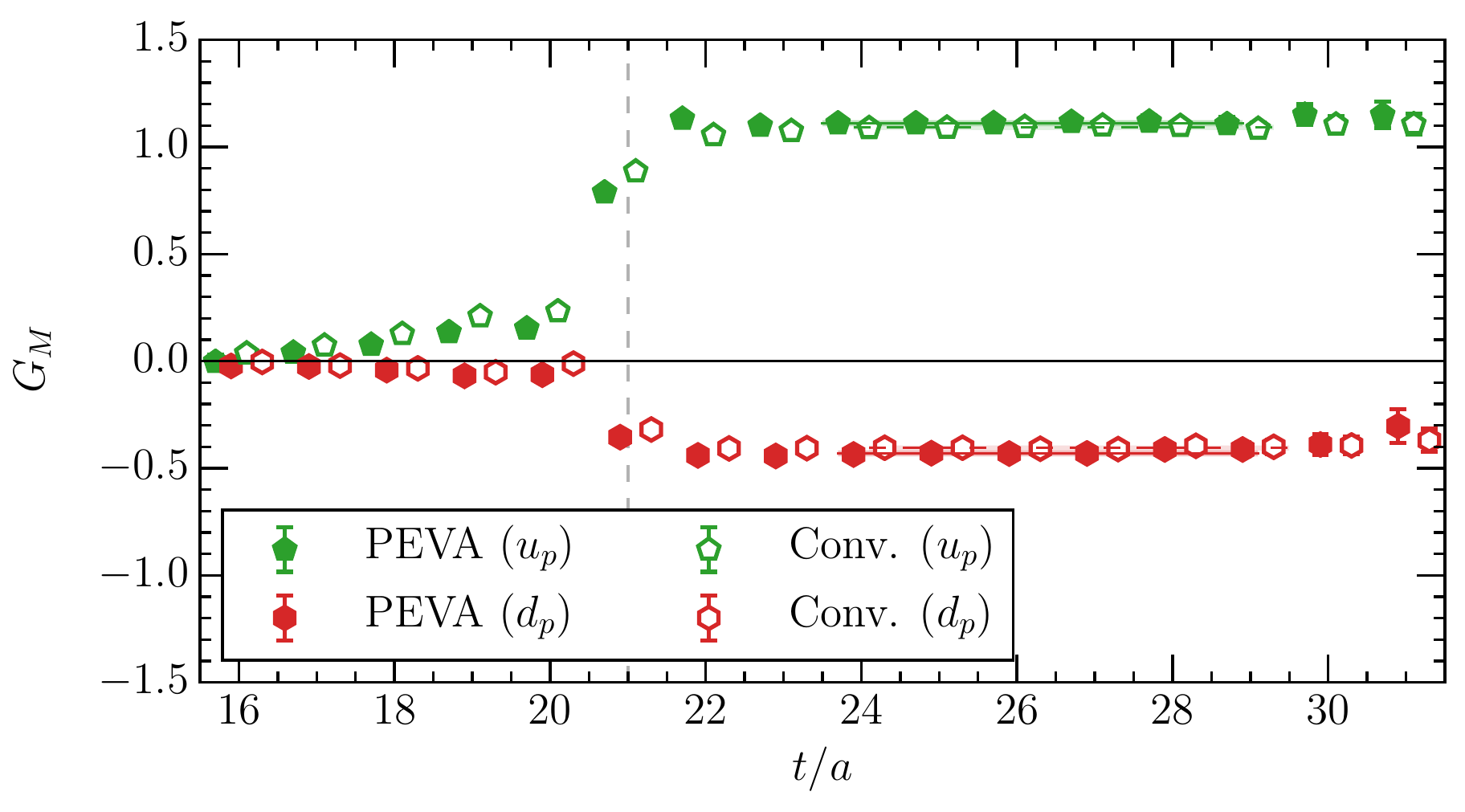}
	\end{center}
	\caption{\label{fig:groundstate:GM}\(G_M(Q^2)\) for the ground state nucleon at
        \(Q^2 = \SI{0.144}{\giga\electronvolt^2}\).
	We plot the conventional analysis with open markers and dashed lines and the
	PEVA analysis with closed markers and solid lines. The
	contributions from the singly represented quark sector
	are plotted with red hexagons, and from the doubly represented quark sector
	with green pentagons, for single quarks of unit charge.}
\end{figure}

Beginning with the ground state, in Fig.~\ref{fig:groundstate:GE} we plot \(G_E(Q^2)\) and in
Fig.~\ref{fig:groundstate:GM} we plot \(G_M(Q^2)\) with respect to sink time, at \(Q^2 = \SI{0.144}{\giga\electronvolt^2}\).
There are only slight differences between the
results extracted by PEVA and the results extracted by a conventional variational analysis in this case.
We believe this is because the opposite parity contaminations are small, and come from heavier states that
are suppressed by Euclidean time evolution.

For both the PEVA and conventional variational analysis, we see very clear and clean plateaus in the form
factors, indicating very good control over excited state contaminations. This supports previous work
demonstrating the utility of variational analysis in calculating baryon matrix elements
\cite{Dragos:2016rtx, Owen:2012ts}. By using such techniques we are able to cleanly isolate precise
values for the Sachs electric and magnetic form factors of the ground state nucleon.

\begin{figure}[t]
	\begin{center}
		\includegraphics[width=0.8\textwidth, trim={0 15mm 0 3mm}]{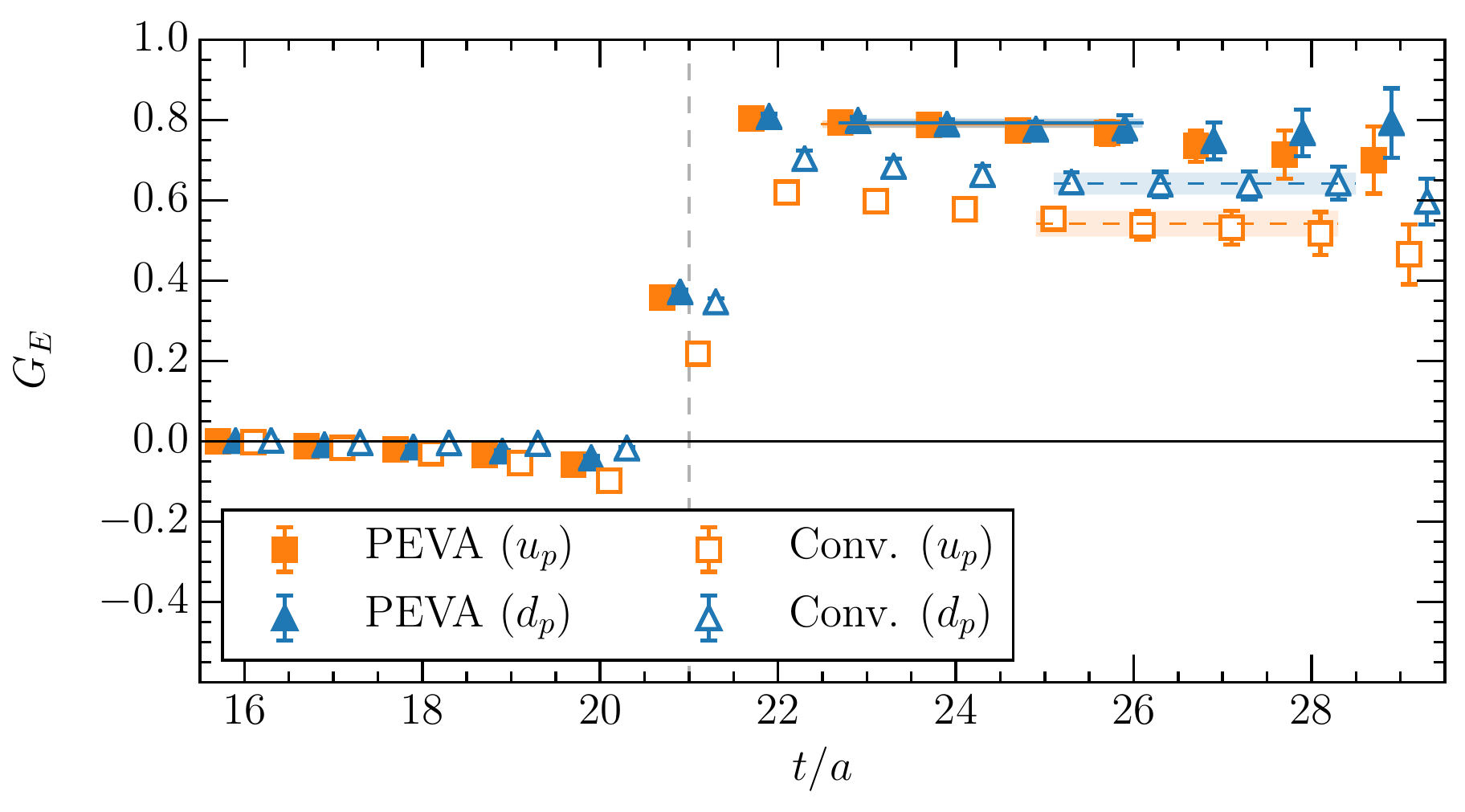}
	\end{center}
	\caption{\label{fig:firstneg:GE}\(G_E(Q^2)\) for the first negative parity excitation of the nucleon at
        \(Q^2 = \SI{0.146}{\giga\electronvolt^2}\).
	We plot the the conventional analysis with open markers and dashed lines, and the new PEVA analysis with closed
	markers and solid lines. We plot the
	contributions from the singly represented quark sector
	with blue triangles, and doubly represented quark sector
	with orange squares, for single quarks of unit charge. The PEVA analysis plateaus at a much earlier time
	with a significantly different value, demonstrating its effectiveness.}
\end{figure}

\begin{figure}[t]
	\begin{center}
		\includegraphics[width=0.8\textwidth, trim={0 15mm 0 3mm}]{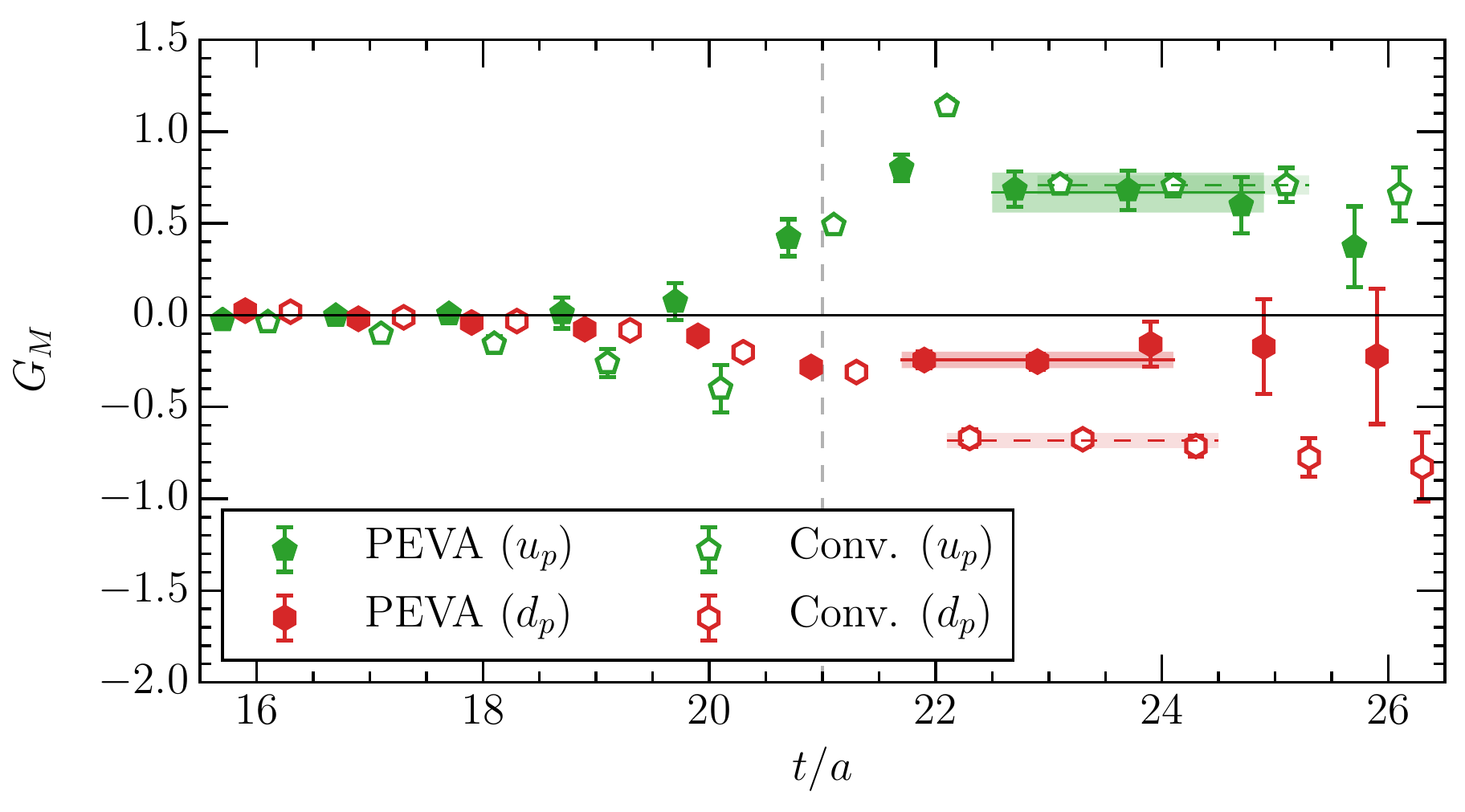}
	\end{center}
	\caption{\label{fig:firstneg:GM}\(G_M(Q^2)\) for the ground state nucleon at
        \(Q^2 = \SI{0.146}{\giga\electronvolt^2}\).
	We plot the conventional analysis with open markers and dashed lines and the
	PEVA analysis with closed markers and solid lines. The
	contributions from the singly represented quark sector
	are plotted with red hexagons, and from the doubly represented quark sector
	with green pentagons, for single quarks of unit charge. The significantly different values
	of \(G_M(Q^2)\) for the singly represented quark sector demonstrate the importance of
	the PEVA technique.}
\end{figure}

Moving on to the first negative parity excited state, in Fig.~\ref{fig:firstneg:GE} we plot \(G_E(Q^2)\)
at \(Q^2 = \SI{0.146}{\giga\electronvolt^2}\) vs the sink time.
We see that the PEVA analysis (closed points) allows us to fit a plateau (shaded bands with solid lines)
at a much earlier time, and with a significantly different value to the 
conventional analysis (open points, dashed lines). This demonstrates the effectiveness of the PEVA technique
at removing opposite parity contaminations. The localised nature of this state is manifest in the large
values for \(G_E(Q^2)\), similar to that for the proton.
We also see similar improvements in \(G_M(Q^2)\), as presented in Fig.~\ref{fig:firstneg:GM}.

\section{Conclusion}

We have demonstrated the effectiveness of the PEVA technique specifically and variational analysis techniques
in general at controlling excited state effects. For the ground state nucleon, conventional variational analysis
techniques are sufficient to provide clean plateaus that allow for the effective extraction of form factors. However,
for excited states, opposite parity contaminations have a clear and
significant effect. The PEVA technique allows us to remove these contaminations and
calculate the form factors of excited nucleons for the first time.

\bibliographystyle{JHEP}
\bibliography{database}

\providecommand{\href}[2]{#2}\begingroup\raggedright\begin{thebibliography}{10}

\bibitem{Kiratidis:2015vpa}
A.~L. Kiratidis, W.~Kamleh, D.~B. Leinweber and B.~J. Owen, \emph{{Lattice
  baryon spectroscopy with multi-particle interpolators}},
  \href{http://dx.doi.org/10.1103/PhysRevD.91.094509}{\emph{Phys. Rev.} {\bf
  D91} (2015) 094509}, [\href{https://arxiv.org/abs/1501.07667}{{\tt
  1501.07667}}].

\bibitem{Mahbub:2013ala}
M.~S. Mahbub, W.~Kamleh, D.~B. Leinweber, P.~J. Moran and A.~G. Williams,
  \emph{{Structure and Flow of the Nucleon Eigenstates in Lattice QCD}},
  \href{http://dx.doi.org/10.1103/PhysRevD.87.094506}{\emph{Phys.Rev.} {\bf
  D87} (2013) 094506}, [\href{https://arxiv.org/abs/1302.2987}{{\tt
  1302.2987}}].

\bibitem{Mahbub:2012ri}
M.~S. Mahbub, W.~Kamleh, D.~B. Leinweber, P.~J. Moran and A.~G. Williams,
  \emph{{Low-lying Odd-parity States of the Nucleon in Lattice QCD}},
  \href{http://dx.doi.org/10.1103/PhysRevD.87.011501}{\emph{Phys. Rev.} {\bf
  D87} (2013) 011501}, [\href{https://arxiv.org/abs/1209.0240}{{\tt
  1209.0240}}].

\bibitem{Mahbub:2010rm}
{\scshape CSSM Lattice} collaboration, M.~S. Mahbub, W.~Kamleh, D.~B.
  Leinweber, P.~J. Moran and A.~G. Williams, \emph{{Roper Resonance in 2+1
  Flavor QCD}},
  \href{http://dx.doi.org/10.1016/j.physletb.2011.12.048}{\emph{Phys. Lett.}
  {\bf B707} (2012) 389--393}, [\href{https://arxiv.org/abs/1011.5724}{{\tt
  1011.5724}}].

\bibitem{Edwards:2012fx}
{\scshape Hadron Spectrum} collaboration, R.~G. Edwards et~al., \emph{{Flavor
  structure of the excited baryon spectra from lattice QCD}},
  \href{http://dx.doi.org/10.1103/PhysRevD.87.054506}{\emph{Phys.Rev.} {\bf
  D87} (2013) 054506}, [\href{https://arxiv.org/abs/1212.5236}{{\tt
  1212.5236}}].

\bibitem{Edwards:2011jj}
R.~G. Edwards, J.~J. Dudek, D.~G. Richards and S.~J. Wallace, \emph{{Excited
  state baryon spectroscopy from lattice QCD}},
  \href{http://dx.doi.org/10.1103/PhysRevD.84.074508}{\emph{Phys. Rev.} {\bf
  D84} (2011) 074508}, [\href{https://arxiv.org/abs/1104.5152}{{\tt
  1104.5152}}].

\bibitem{Lang:2016hnn}
C.~B. Lang, L.~Leskovec, M.~Padmanath and S.~Prelovsek, \emph{{Pion-nucleon
  scattering in the Roper channel from lattice QCD}},
  \href{https://arxiv.org/abs/1610.01422}{{\tt 1610.01422}}.

\bibitem{Lang:2012db}
C.~Lang and V.~Verduci, \emph{{Scattering in the ${\pi}{N}$ negative parity
  channel in lattice QCD}},
  \href{http://dx.doi.org/10.1103/PhysRevD.87.054502}{\emph{Phys.Rev.} {\bf
  D87} (2013) 054502}, [\href{https://arxiv.org/abs/1212.5055}{{\tt
  1212.5055}}].

\bibitem{Alexandrou:2014mka}
C.~Alexandrou, T.~Leontiou, C.~N. Papanicolas and E.~Stiliaris, \emph{{Novel
  analysis method for excited states in lattice QCD: The nucleon case}},
  \href{http://dx.doi.org/10.1103/PhysRevD.91.014506}{\emph{Phys. Rev.} {\bf
  D91} (2015) 014506}, [\href{https://arxiv.org/abs/1411.6765}{{\tt
  1411.6765}}].

\bibitem{Menadue:2013kfi}
F.~M. Stokes et~al., \emph{{Parity-expanded variational analysis for nonzero
  momentum}}, \href{http://dx.doi.org/10.1103/PhysRevD.92.114506}{\emph{Phys.
  Rev.} {\bf D92} (2015) 114506}, [\href{https://arxiv.org/abs/1302.4152}{{\tt
  1302.4152}}].

\bibitem{Lee:1998cx}
F.~X. Lee and D.~B. Leinweber, \emph{{Negative parity baryon spectroscopy}},
  \href{http://dx.doi.org/10.1016/S0920-5632(99)85041-5}{\emph{Nucl. Phys.
  Proc. Suppl.} {\bf 73} (1999) 258--260},
  [\href{https://arxiv.org/abs/hep-lat/9809095}{{\tt hep-lat/9809095}}].

\bibitem{Martinelli:1990ny}
G.~Martinelli, C.~T. Sachrajda and A.~Vladikas, \emph{{A Study of 'improvement'
  in lattice QCD}},
  \href{http://dx.doi.org/10.1016/0550-3213(91)90538-9}{\emph{Nucl. Phys.} {\bf
  B358} (1991) 212--227}.

\bibitem{Boinepalli:2006xd}
S.~Boinepalli et~al., \emph{{Precision electromagnetic structure of octet
  baryons in the chiral regime}},
  \href{http://dx.doi.org/10.1103/PhysRevD.74.093005}{\emph{Phys. Rev.} {\bf
  D74} (2006) 093005}, [\href{https://arxiv.org/abs/hep-lat/0604022}{{\tt
  hep-lat/0604022}}].

\bibitem{stokes:formfactors}
F.~M. Stokes, W.~Kamleh, D.~B. Leinweber and B.~J. Owen{\emph{{,}} {in
  preparation}}.

\bibitem{Aoki:2008sm}
{\scshape PACS-CS} collaboration, S.~Aoki et~al., \emph{{2+1 Flavor Lattice QCD
  toward the Physical Point}},
  \href{http://dx.doi.org/10.1103/PhysRevD.79.034503}{\emph{Phys. Rev.} {\bf
  D79} (2009) 034503}, [\href{https://arxiv.org/abs/0807.1661}{{\tt
  0807.1661}}].

\bibitem{Beckett:2009cb}
M.~G. Beckett, B.~Joo, C.~M. Maynard, D.~Pleiter, O.~Tatebe and T.~Yoshie,
  \emph{{Building the International Lattice Data Grid}},
  \href{http://dx.doi.org/10.1016/j.cpc.2011.01.027}{\emph{Comput. Phys.
  Commun.} {\bf 182} (2011) 1208--1214},
  [\href{https://arxiv.org/abs/0910.1692}{{\tt 0910.1692}}].

\bibitem{Dragos:2016rtx}
J.~Dragos et~al., \emph{{Nucleon matrix elements using the variational method
  in lattice QCD}},
  \href{http://dx.doi.org/10.1103/PhysRevD.94.074505}{\emph{Phys. Rev.} {\bf
  D94} (2016) 074505}, [\href{https://arxiv.org/abs/1606.03195}{{\tt
  1606.03195}}].

\bibitem{Owen:2012ts}
B.~J. Owen, J.~Dragos, W.~Kamleh, D.~B. Leinweber, M.~S. Mahbub, B.~J. Menadue
  et~al., \emph{{Variational Approach to the Calculation of gA}},
  \href{http://dx.doi.org/10.1016/j.physletb.2013.04.063}{\emph{Phys. Lett.}
  {\bf B723} (2013) 217--223}, [\href{https://arxiv.org/abs/1212.4668}{{\tt
  1212.4668}}].

\end{thebibliography}\endgroup

\end{document}